\documentclass[prl, twocolumn, superscriptaddress, longbibliography]{revtex4-1}

\usepackage{amssymb}
\usepackage{amsmath}
\usepackage{graphicx}
\usepackage{amsfonts}
\usepackage[usenames, dvipsnames]{xcolor}
\usepackage{bm}
\usepackage{bbm}
\usepackage{color}
\usepackage{hyperref}
\hypersetup{pdfnewwindow=true,
	colorlinks=true,
	linkcolor=blue,
	citecolor=blue,
	urlcolor=blue}
\usepackage{soul}
\usepackage[caption=false, justification=centerlast]{subfig}
\usepackage{psfrag}
\usepackage{textcomp}
\usepackage{braket}

\begin{document}
\title{Catalytic leverage of correlations and mitigation of dissipation in
Information erasure}
\author{I. Henao$^{1}$}
\email{ivan.henao@mail.huji.ac.il}

\author{R. Uzdin$^{1}$}
\email{raam@mail.huji.ac.il}

\affiliation{$^{1}$Fritz Haber Research Center for Molecular Dynamics,Institute
of Chemistry, The Hebrew University of Jerusalem, Jerusalem 9190401,
Israel}
\begin{abstract}
Correlations are a valuable resource for quantum information processing and
quantum thermodynamics. However, the preparation of some
correlated states can carry a substantial cost that should be compared
against its value. We show that classical correlations can be catalytically
exploited, which enables to mitigate heat and entropy dissipation
in information erasure. These correlations are naturally generated
by the erasure process, and thus can be considered free. Although we also show that 
maximum erasure with minimum dissipation and no correlations is theoretically possible, 
catalysts are always useful in practical erasure settings, where correlations are 
expected to take place. 

\end{abstract}
\maketitle
Catalysts are a powerful asset in quantum thermodynamics \cite{brandao2015second,lostaglio2015stochastic,ng2015limits,muller2018correlating,Boes2020passingfluctuation,henao2021catalytic,PhysRevX11011061,shiraishi2021quantum} and related quantum information processing tasks \cite{aaberg2014catalytic,boes2019neumann,PhysRevLett.127.080502}. Similarly to chemical catalysts, quantum catalysts are capable to assist physical processes without being modified. In quantum thermodynamics, catalysts have profound consequences on the ability to manipulate a physical system. Catalysts lift stringent thermodynamic restrictions at the nanoscale \cite{brandao2015second,lostaglio2015stochastic,muller2018correlating,shiraishi2021quantum}, significantly increasing the set of transformations
that are possible in the presence of a thermal environment. Moreover,
this potential can be released without requiring fine-tuned catalysts
\cite{PhysRevX11011061}. 

Catalysts have also proven fruitful in situations where infinitely large environments
are unaccessible. For example, a single quantum
system in a non-thermal passive state can deliver work if coupled
to a sufficiently large catalyst \cite{sparaciari2017energetic}. Cooling
is another important thermodynamic task that can benefit from catalysts.
Given a too small and/or too hot environment, cooling can be catalytically
activated or enhanced using a catalyst of finite dimension \cite{henao2021catalytic}. 

In this Letter, we study how catalysts help to mitigate the thermodynamic impact of information erasure on the environment. Information erasure can be associated with any transformation that takes an initial
quantum state $\rho_{s}$ of a system $s$, and outputs a final state
$\rho'_{s}$ having less entropy than $\rho_{s}$. A  widely
used quantifier of entropy for quantum states is the von Neumann entropy
$S(\rho_{s})$, which generalizes the classical Shannon entropy. Information
erasure is at the heart of the connection between information theory
and thermodynamics, linking a reduction of entropy $-\Delta S_{s}=S(\rho_{s})-S(\rho'_{s})>0$
with a minimum amount of heat dissipated into the environment. Such
a thermodynamic cost is the core of Landauer's principle \cite{landauer1961irreversibility,vinjanampathy2016quantum}.
Here, we show that when information is erased using a \textit{finite}
environment, correlations between the system and the environment can
be catalytically exploited to reduce heat and entropy waste. 
\begin{figure}

\centering{}\includegraphics[scale=1.02]{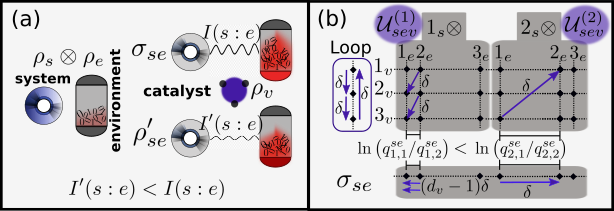}\caption{Catalytic mitigation of dissipation in information erasure. (a) Initially,
information is erased from a memory (the \textquotedblleft system\textquotedblright{}
$s$) in a mixed state $\rho_{s}$, by coupling it to a finite environment
$e$ in a (potentially thermal) state $\rho_{e}$. The resulting state
$\sigma_{se}$ is in general correlated and heat and entropy are dissipated
into $e$. By using a \textit{finite} catalyst $v$, mutual information
$I(s:e)$ associated with correlations in $\sigma_{se}$ can
be consumed to reduce both forms of dissipation, while keeping the
system state $\sigma_{s}=\textrm{Tr}_{e}(\sigma_{se})$ and the catalyst
state $\rho_{v}$ unchanged. (b) $\textrm{ln}(p^{se})\times\textrm{ln}(p^{v})$
diagram of a state $\sigma_{se}\otimes\rho_{v}$, where $\sigma_{se}$
is a state of a qubit system and a qutrit environment, and $\rho_{v}$
is the state of a qutrit catalyst. A catalytic
transformation that mitigates dissipation is generated by a permutation
$\mathcal{U}_{sev}^{(1)}\oplus\mathcal{U}_{sev}^{(2)}$, where $\mathcal{U}_{sev}^{(1)}$
acts on the subspace $|1_{s}\rangle\otimes\{|j_{e}\rangle\}$ (upper
left box) and $\mathcal{U}_{sev}^{(2)}$ acts on the subspace $|2_{s}\rangle\otimes\{|j_{e}\rangle\}$
(upper right box). The purple arrows depict population transfers associated
to this transformation. The \textquotedblleft Loop\textquotedblright{}
and the lower box illustrate how these transfers act on the catalyst
and the compound $se$, respectively. }
\label{Fig. 1}
\end{figure}

The role of quantum and classical correlations has been extensively studied in thermodynamics \cite{bera2017generalized,sapienza2019correlations}. This includes diverse tasks such as work extraction \cite{funo2013thermodynamic,PhysRevLett.111.240401,perarnau2015extractable,francica2017daemonic,PhysRevLett.121.120602},  information erasure \cite{del2011thermodynamic}, the charging \cite{binder2015quantacell,ferraro2018high} or discharging \cite{PhysRevE.87.042123,andolina2019extractable} of quantum batteries, and heat flow reversals that
are forbidden in uncorrelated systems  \cite{jennings2010entanglement,henao2018role}.
In the catalytic scenario, correlations lead to enhanced  catalytic transformations \cite{muller2018correlating,shiraishi2021quantum,lostaglio2015stochastic}. For example, stringent thermodynamic constraints in small
quantum systems \cite{brandao2015second} are relaxed to the traditional second (weaker) law \cite{muller2018correlating} once the catalyst is allowed to get correlated with the system. Importantly, these correlations are a byproduct of the transformation, as opposed to initial correlations whose creation involves a thermodynamic cost \cite{huber2015thermodynamic,PhysRevE.91.032118}. 

In this work, we demonstrate that a catalyst can mitigate dissipation
for information erasure, through the correlations naturally generated by
the erasure transformation (see Fig. \ref{Fig. 1}(a)). We also identify a special class of 
optimal erasure transformations that give rise to uncorrelated states. While these transformations 
are interesting from a theoretical perspective, it is worth stressing that whenever correlations 
arise catalysts play a mitigating role. Therefore, realistic erasure processes can always benefit from
catalysts.

We start by describing the setting for standard erasure, i.e. without
including a catalyst. Initially, a finite environment $e$ in a state
$\rho_{e}$ is used to reduce the entropy of a system $s$ in the
initial state $\rho_{s}$. This is achieved through a global unitary
evolution $U_{se}$ that acts jointly on the uncorrelated state $\rho_{se}=\rho_{s}\otimes\rho_{e}$.
The correlations of the resulting state $\sigma_{se}=U_{se}(\rho_{s}\otimes\rho_{e})U_{se}^{\dagger}$
can be quantified by the mutual information $I(s:e)\equiv\Delta S_{s}+\Delta S_{e}$,
where $\Delta S_{e}=S(\sigma_{e})-S(\rho_{e})$ is the entropy change
of the environment, and $\sigma_{e}=\textrm{Tr}_{s}(\sigma_{se})$.\textcolor{green}{{} }

For any erasure transformation $\rho_{s}\rightarrow\sigma_{s}=\textrm{Tr}_{e}(\sigma_{se})$
(such that $S(\rho_{s})>S(\sigma_{s})$), the positivity of $I(s:e)$
implies that entropy $\Delta S_{e}>0$ is dissipated into the environment.
The link with heat dissipation can be established from first principlies
via the equality \cite{reeb2014improved}

\begin{equation}
\beta_{e}Q_{e}=-\Delta S_{s}+I(s:e)+S(\sigma_{e}||\rho_{e}),\label{eq:1 Landauer's principle}
\end{equation}
which describes the heat dissipated $Q_{e}=\textrm{Tr}\left[H_{e}(\sigma_{e}-\rho_{e})\right]$
into an environment that starts in the \textit{thermal} state $\rho_{e}=\frac{e^{-\beta_{e}H_{e}}}{\textrm{Tr}(e^{-\beta_{e}H_{e}})}$.
Here, $H_{e}$ is the environment Hamiltonian and $\beta_{e}=1/T_{e}$
is the inverse of the corresponding temperature $T_{e}$. Moreover,
$S(\sigma_{e}||\rho_{e})=-\textrm{Tr}(\sigma_{e}\textrm{ln}\rho_{e})-S(\sigma_{e})$
is the quantum relative entropy between $\sigma_{e}$
and $\rho_{e}$ \cite{sagawa2013second}. 

Since $S(\sigma_{e}||\rho_{e})\geq0$ for any pair $\sigma_{e}$ and
$\rho_{e}$, it immediately follows from Eq. (\ref{eq:1 Landauer's principle})
that $\beta_{e}Q_{e}\geq-\Delta S_{s}$, which represents the standard
inequality form of Landauer's principle. Unless the environment is
\textit{infinite}, the contribution $I(s:e)+S(\sigma_{e}||\rho_{e})$
to $\beta_{e}Q_{e}$ is not negligible \cite{reeb2014improved}.
The primary goal that we set for the catalyst is to reduce $Q_{e}$
to a value $Q'_{e}<Q{}_{e}$, while keeping the state $\sigma_{s}$
unchanged. Hence, we have a composition of transformations $\rho_{se}\rightarrow\sigma_{se}\otimes\rho_{v}\rightarrow\rho'_{sev}$,
where $\rho'_{sev}=U_{sev}\left(\sigma_{se}\otimes\rho_{v}\right)U_{sev}^{\dagger}$
and $\rho_{v}$ is a suitable catalyst state such that $Q'_{e}=\textrm{Tr}\left[H_{e}(\rho'_{sev}-\rho_{sev})\right]=\textrm{Tr}\left[H_{e}(\rho'_{e}-\rho_{e})\right]<Q_{e}$.
Apart from the requirement $\rho'_{s}=\textrm{Tr}_{ev}(\rho'_{sev})=\sigma_{s}$,
the transformation is catalytic if and only if $\rho'_{v}=\textrm{Tr}_{se}(\rho'_{sev})=\rho_{v}$.
We also remark that since $U_{sev}$ is unitary, the catalyst is the
only additional system assisting information erasure. 

The key property for our characterization of transformations that
mitigate heat and entropy is majorization \cite {marshall1979inequalities,nielsen2002introduction}. 
The formal definition of this property is given in \footnote{see Supplemental Material}. For our purposes, it suffices to
know that if $\rho'_{e}$ majorizes $\sigma_{e}$, symbolically written
as $\rho'_{e}\succ\sigma_{e}$, then $Q'_{e}\leq Q_{e}$ (cf. Lemma 1
in \cite{Note1}). Not only that, but also $S(\rho'_{e})\leq S(\sigma_{e})$
due to the Schur-concavity of the von-Neumann entropy \cite{nielsen2002introduction}.
Therefore, Theorem 1 implies that in any erasure transformation that
yields a correlated state $\sigma_{se}$ it is possible to catalytically
mitigate both heat and entropy dissipation. 

\textbf{Theorem 1}. There exists a catalyst state $\rho_{v}$ and
a transformation $\sigma_{se}\otimes\rho_{v}\rightarrow\rho'_{sev}=U_{sev}\left(\sigma_{se}\otimes\rho_{v}\right)U_{sev}^{\dagger}$,
such that $\rho'_{e}=\textrm{Tr}_{sv}(\rho'_{sev})\succ\sigma_{e}$,
$\textrm{Tr}_{ev}(\rho'_{sev})=\sigma_{s}=\textrm{Tr}_{e}(\sigma_{se})$,
and $\textrm{Tr}_{se}(\rho'_{sev})=\rho_{v}$, if and only if $\sigma_{se}$
is a correlated state. 

\textit{Proof of ``Only if'' implication}.   We shall resort to the already noted implication $\rho'_{e}\succ\sigma_{e}\Rightarrow S(\rho'_{e})\leq S(\sigma_{e})$. We use primed symbols
to denote quantities associated with $\rho'_{sev}$ or variations
associated to the transformation $\sigma_{se}\otimes\rho_{v}\rightarrow\rho'_{sev}$.
Since $\sigma_{se}\otimes\rho_{v}$ is an uncorrelated state, the
mutual information between $se$ and $v$ can only increase. Namely,
$I'(se:v)=\Delta'S_{se}+\Delta'S_{v}=\Delta'S_{se}\geq0$. Moreover,
the condition $\Delta'S_{s}=0$ implies that $\Delta'I(s:e)=\Delta'S_{s}+\Delta'S_{e}-\Delta'S_{se}=\Delta'S_{e}-\Delta'S_{se}$.
Since $\Delta'S_{se}\geq0$, a reduction of the environment entropy
$\Delta'S_{e}=S(\rho'_{e})-S(\sigma_{e})\leq0$ implies $\Delta'I(s:e)=I'(s:e)-I(s:e)<0$,
which is possible only if $I(s:e)>0$. 

\textit{Proof of ``If'' implication}. In what follows we will refer to some technical results, presented in \cite{Note1}.
Let $\{|i_{s}\rangle\}_{i=1}^{d_{s}}$ and $\{|j_{e}\rangle\}_{j=1}^{d_{e}}$
denote respectively orthonormal bases of $\mathcal{H}_{s}$ and $\mathcal{H}_{e}$,
where $\mathcal{H}_{x}$ is the Hilbert space of system $x$ and $d_{x}$
is its dimension. Lemma 2 of \cite{Note1} states that
if a joint state $\sigma_{se}$ with eigendecomposition
$\sigma_{se}=\sum_{i=1}^{d_{s}}\sum_{j=1}^{d_{e}}q_{i,j}^{se}|i_{s}j_{e}\rangle\langle i_{s}j_{e}|$
is correlated, then $\frac{q_{I,J}^{se}}{q_{I,J'}^{se}}>\frac{q_{I',J}^{se}}{q_{I',J'}^{se}}\geq1$,
for some tuple $(I,I',J,J')$. This inequality is key for the following
construction. We also write the eigendecomposition of $\rho_{v}$
as $\rho_{v}=\sum_{k=1}^{d_{v}}p_{k}^{v}|k_{v}\rangle\langle k_{v}|$,
and set (without loss of generality) $p_{k}^{v}\geq p_{k+1}^{v}$
for $1\leq k\leq d_{v}-1$.\textcolor{green}{{} }

If $\sigma_{se}$ satisfies $q_{I,J}^{se}/q_{I,J'}^{se}>q_{I',J}^{se}/q_{I',J'}^{se}\geq1$,
we can also choose catalyst eigenvalues that fulfill the inequalities 

\begin{equation}
\frac{q_{I,J}^{se}}{q_{I,J'}^{se}}>\frac{p_{1}^{v}}{p_{d_{v}}^{v}}>\frac{p_{k}^{v}}{p_{k+1}^{v}}>\frac{q_{I',J}^{se}}{q_{I',J'}^{se}}\geq1,\:1\leq k\leq d_{v}-1.\label{eq:2 inequal for catalytic transformation}
\end{equation}
Let us now define $\mathcal{U}_{i_{s}j_{e}k_{v}\leftrightarrow i'_{s}j'_{e}k'_{v}}$
as a permutation that swaps the global eigenstates $|i_{s}j_{e}k_{v}\rangle$
and $|i'_{s}j'_{e}k'_{v}\rangle$, while acting as the identity on
any other eigenstate of $sev$. Any composition of such permutations
produces a total state that is diagonal in the eigenbasis of $\sigma_{se}\otimes\rho_{v}$,
as well as local states that are diagonal in the original eigenbases
(e.g. the final state of $e$ is diagonal in the eigenbasis of $\sigma_{e}$).
Therefore, this transformation only changes the global and local eigenvalues.
Keeping in mind the rightmost inequalities in Eq. (\ref{eq:2 inequal for catalytic transformation}),
a permutation $\mathcal{U}_{I'_{s}J'_{e}k_{v}\leftrightarrow I'_{s}J_{e}(k+1)_{v}}$
increases the eigenvalue of $|J_{e}\rangle$ and reduces that of $|J'_{e}\rangle$
by the same amount $\delta_{k}\equiv q_{I',J'}^{se}p_{k}^{v}-q_{I',J}^{se}p_{k+1}^{v}$.
These variations of eigenvalues also describe ``population transfers''
between global eigenstates and in the local eigenbases of the different
systems $x=s,e,v$, which we will often refer to in the following. 

Suppose now that the eigenvalues $\{p_{k}^{v}\}$ are chosen in such
a way that $\delta_{k}=\delta_{k+1}\equiv\delta$ for any $1\leq k\leq d_{v}-1$.
As proven in \cite{Note1} (Lemma 3),
this is always possible. In this way, a direct sum of permutations
$\mathcal{U}_{sev}^{(I')}\equiv\bigoplus_{k=1}^{d_{v}-1}\mathcal{U}_{I'_{s}J'_{e}k_{v}\leftrightarrow I'_{s}J_{e}(k+1)_{v}}$
transfers total population $\sum_{k=1}^{d_{v}-1}\delta_{k}=(d_{v}-1)\delta$
from $|J'_{e}\rangle$ to $|J_{e}\rangle$. Moreover, each permutation
$\mathcal{U}_{I'_{s}J'_{e}k_{v}\leftrightarrow I'_{s}J_{e}(k+1)_{v}}$
transfers population $\delta$ from $|k_{v}\rangle$ to $|(k+1)_{v}\rangle$.
The net effect of $\mathcal{U}_{sev}^{(I')}$ on $v$ is thus to increase
the eigenvalue of $|d_{v}\rangle$ and to reduce the eigenvalue of
$|1_{v}\rangle$ by the same amount $\delta$. 

On the other hand, the leftmost inequality in (\ref{eq:2 inequal for catalytic transformation})
guarantees that a permutation $\mathcal{U}_{sev}^{(I)}\equiv\mathcal{U}_{I_{s}J'_{e}1_{v}\leftrightarrow I_{s}J_{e}d_{v}}$
transfers population $\delta_{d_{v}}\equiv q_{I,J}^{se}p_{d_{v}}^{v}-q_{I,J'}^{se}p_{1}^{v}$
from $|d_{v}\rangle$ to $|1_{v}\rangle$, and from $|J_{e}\rangle$
to $|J'_{e}\rangle$. If $\delta_{d_{v}}=\delta$ and we denote the
eigenvalues of $\sigma_{e}=\textrm{Tr}_{s}\sigma_{se}$ as $\{q_{j}^{e}\}$,
it follows that $U_{sev}=\mathcal{U}_{sev}^{(I')}\oplus\mathcal{U}_{sev}^{(I)}$
satisfies $\rho'_{v}=\textrm{Tr}_{se}\rho'_{sev}=\rho_{v}$ and $\rho'_{e}=\textrm{Tr}_{sv}\rho'_{sev}=\sum_{j}p_{j}^{\prime e}|j_{e}\rangle\langle j_{e}|$,
where $p_{J}^{\prime e}=q_{J}^{e}+(d_{v}-2)\delta$, $p_{J'}^{\prime e}=q_{J'}^{e}-(d_{v}-2)\delta$,
and $p_{j}^{\prime e}=q_{j}^{e}$ for $j\neq J,J'$. For
$q_{J}^{e}=\sum_{i=1}^{d_{s}}q_{i,J}^{se}>q_{J'}^{e}$ and $d_{v}\geq3$,
Lemma 4 of \cite{Note1} implies that
$\rho'_{e}$ majorizes $\sigma_{e}$. 

Finally, notice that $\mathcal{U}_{sev}^{(I')}=|I'_{s}\rangle\langle I'_{s}|\otimes\bigoplus_{k=1}^{d_{v}-1}\mathcal{U}_{J'_{e}k_{v}\leftrightarrow J_{e}(k+1)_{v}}$
and $\mathcal{U}_{sev}^{(I)}=|I_{s}\rangle\langle I_{s}|\otimes\mathcal{U}_{J'_{e}1_{v}\leftrightarrow J_{e}d_{v}}$,
where $\mathcal{U}_{j_{e}k_{v}\leftrightarrow j'_{e}k'_{v}}$ is a
permutation that exchanges $|j_{e}k_{v}\rangle$ and $|j'_{e}k'_{v}\rangle$.
Therefore, $U_{sev}$ is a controlled unitary where the system acts
as control. This ensures that $\rho'_{s}=\textrm{Tr}_{ev}\rho'_{sev}=\sigma_{s}$
and that erasure $\rho_{s}\rightarrow\sigma_{s}$ is not affected
by $U_{sev}$. 

To complement the sufficiency proof, Fig. \ref{Fig. 1}(b) illustrates the case
of a state $\sigma_{se}\otimes\rho_{v}$ where $s$ is a qubit, and
$e$ and $v$ are qutrits \footnote{If the system is a qubit, a qutrit environment is the smallest environment for which a catalyst can offer an advantage by exploiting the correlations of $\sigma_{se}$. As implied by Theorems 2 and 3, if $d_{e}=2$ a simple swap between the two qubits implements an optimal erasure transformation, which does not admit any improvement regarding reduction of dissipation (see also the example in the last section of \cite{Note1})}.
The state $\sigma_{se}\otimes\rho_{v}$ is depicted using a $\textrm{ln}(p^{se})\times\textrm{ln}(p^{v})$
diagram \cite{henao2021catalytic}. In this diagram ``columns'' $i_{s}j_{e}=i_{s}\otimes j_{e}$
(vertical lines) represent the eigenstates $|i_{s}j_{e}\rangle$,
and ``rows'' $k_{v}$ (horizontal lines) represent the eigenstates
$|k_{v}\rangle$. The columns are arranged from left to right, placing
first the column with largest eigenvalue, then the column with the
next largest eigenvalue, and so on. Similarly, for eigenvalues $p_{k}^{v}\geq p_{k'}^{v}$
the row $k_{v}$ is located above the row $k'_{v}$, and the distance between these rows reads $\textrm{ln}(p_{k}^{v}/p_{k'}^{v})$. Likewise, two columns with eigenvalues $q_{i,j}^{se}$ and $q_{i',j'}^{se}\leq q_{i,j}^{se}$ are separated by a distance $\textrm{ln}(q_{i,j}^{se}/q_{i',j'}^{se})$. Each intersection
between $i_{s}j_{e}$ and $k_{v}$ stands for the global eigenstate
$|i_{s}j_{e}k_{v}\rangle$. 

For the state $\sigma_{se}\otimes\rho_{v}$ in Fig. \ref{Fig. 1}(b) we have the
tuple $(I,I',J,J')=(2,1,1,2)$, which also characterizes the permutation
$\mathcal{U}_{sev}^{(1)}\oplus\mathcal{U}_{sev}^{(2)}$. The purple
arrows in the left (right) gray box of Fig. \ref{Fig. 1}(b) depict the population
transfers between eigenstates of $\sigma_{se}\otimes\rho_{v}$, generated
by the permutation $\mathcal{U}_{sev}^{(1)}$ ($\mathcal{U}_{sev}^{(2)}$).
The lower gray box shows how these transfers occur between eigenstates
of $\sigma_{se}$. From the point of view of the catalyst, such transfers
compose the loop shown at the left hand side. In this way, each eigenstate
of $v$ has an incoming and an outogoing transfer that cancel each
other, keeping the state of the catalyst unchanged. 

\textbf{\textit{Information erasure with minimum dissipation of heat 
and entropy}}. For two states $\rho_{s}$ and $\rho_{e}$
such that $S(\rho_{e})<S(\rho_{s})$, $\textrm{rank}(\rho_{e})\leq d_{s}$,
and $\textrm{rank}(\rho_{s})\leq d_{e}$, a simple swap between $s$
and $e$ generates an uncorrelated state $\sigma_{s}\otimes\sigma_{e}$,
where $\sigma_{s}=\rho_{e}$ and $\sigma_{e}=\rho_{s}$. Such transformations
minimize entropy dissipation by construction, since $\Delta S_{e}=-\Delta S_{s}$.
However, they are rather limiting in two aspects. First, one may argue
that in practice there is always a non-zero probability for a system
to occupy any of its accessible levels. This corresponds to full-rank
states, which satisfy $\textrm{rank}(\rho_{x})=d_{x}$ for $x=s,e$.
And secondly, for such states the condition $\textrm{rank}(\rho_{e})\leq d_{s}$
implies $d_{e}\leq d_{s}$, which excludes environments that are larger
than the system. 

These limitations motivate us to introduce Theorem 2, where we characterize
erasure transformations for full-rank states $\rho_{x=s,e}$ that
admit minimum entropy dissipation. Notoriously, such transformations
also perform \textit{maximum erasure}. This means that the resulting
state $\sigma_{s}$ has minimum entropy over any system state $\textrm{Tr}_{e}\left[U_{se}(\rho_{s}\otimes\rho_{e})U_{se}^{\dagger}\right]$,
unitarily obtained from $\rho_{s}\otimes\rho_{e}$. Hence, maximum
erasure and minimum entropy dissipation are not incompatible. The
proof of Theorem 2 is given in \cite{Note1}. 

\textbf{Theorem 2}. Consider full-rank states $\rho_{x=s,e}=\sum_{i=1}^{d_{x}}p_{i}^{x}|i_{x}\rangle\langle i_{x}|$,
with non-increasing eigenvalues $p_{i}^{x}\geq p_{i+1}^{x}$ for $1\leq i\leq d_{x}-1$.
Suppose that $\frac{p_{j}^{e}}{p_{j+1}^{e}}\geq\frac{p_{1}^{s}}{p_{d_{s}}^{s}}$
for $1\leq j\leq d_{e}-1$ and that any of the following conditions
holds for positive integers $m,\lambda\geq1$: (i) $d_{e}=md_{s}$
and $r_{e,j}\equiv\frac{p_{j}^{e}}{p_{j+1}^{e}}$ is periodic with
period $m$, i.e. $r_{e,j+\lambda m}=r_{e,j}$, or (ii) $d_{s}=md_{e}$
and $r_{s,i}\equiv\frac{p_{i}^{s}}{p_{i+1}^{s}}$ is periodic with
period $d_{e}$, i.e. $r_{s,i+\lambda d_{e}}=r_{s,i}$. Then there
exists a permutation $\mathcal{V}_{se}$ that maps eigenstates of
$\rho_{s}\otimes\rho_{e}$ into eigenstates of $\rho_{s}\otimes\rho_{e}$,
such that $\mathcal{V}_{se}(\rho_{s}\otimes\rho_{e})\mathcal{V}_{se}^{\dagger}=\sigma_{s}\otimes\sigma_{e}$
and $S(\sigma_{s})\leq S\left(\textrm{Tr}_{e}\left[U_{se}(\rho_{s}\otimes\rho_{e})U_{se}^{\dagger}\right]\right)$
for any unitary $U_{se}$.

The cases (i) and (ii) in Theorem 2 are sufficiently general to capture
many physically relevant situations, including environments larger
{[}case (i){]} or smaller {[}case (ii){]} than the system. For example,
(i) is satisfied if $se$ consists of a qubit system and an environment
of even dimension in a thermal state $\rho_{e}$, with Hamiltonian
$H_{e}=\sum_{j}j\omega_{e}|j_{e}\rangle\langle j_{e}|$. Note that
in this particular case $r_{e,j}=e^{\beta_{e}\omega_{e}}$ is independent
of $j$. If $p_{1}^{s}/p_{2}^{s}<e^{\beta_{e}\omega_{e}}$, we have
all the ingredients for the existence of $\mathcal{V}_{se}$, according
to Theorem 2. On the other hand, condition (ii) only poses constraints
on the system state $\rho_{s}$. 

Theorem 2 serves as a starting point for \textit{optimal erasure}
transformations, described in Theorem 3. Namely, transformations that
not only perform maximum erasure with minimum entropy dissipation,
but also minimize the dissipated heat. We have already mentioned that
for $e$ finite $Q_{e}$ is strictly above Landauer's limit.
The tightest (lower) bound on $Q_{e}$ can be obtained by resorting
to the fact that thermal states minimize the energy for a fixed entropy
\cite{pusz1978passive,lenard1978thermodynamical}. Therefore, in
any erasure transformation $\rho_{s}\otimes\rho_{e}\rightarrow\sigma_{se}$
that increases the environment entropy by $\Delta S_{e}$, we have
that $Q_{e}\geq\textrm{Tr}\left\{ H_{e}\left(\rho_{e}^{th}[S(\rho_{e})+\Delta S_{e}]-\rho_{e}\right)\right\} $,
where $\rho_{e}^{th}[S(\rho_{e})+\Delta S_{e}]$ denotes the thermal
state with entropy $S(\rho_{e})+\Delta S_{e}$. Since the energy of
thermal states is an increasing function of its entropy, $Q_{e}$
attains its minimum when $\Delta S_{e}$ is minimized. That is, for
minimum entropy dissipation $\Delta S_{e}=-\Delta S_{s}$. 

\textbf{Theorem 3}. Let $\rho_{se}=\rho_{s}\otimes\rho_{e}$ be a
state with marginals $\rho_{x=s,e}$ that obey Theorem 2, and let
$\rho_{se}=\sum_{j=1}^{d_{s}d_{e}}p_{j}^{se}|j_{se}\rangle\langle j_{se}|$
denote its eigendecomposition, with non-increasing eigenvalues $p_{j}^{se}\geq p_{j+1}^{se}$.
If $\rho_{e}$ is a thermal state, after the permutation $\mathcal{V}_{se}$
in Theorem 2 minimum heat dissipation is possible if: (i) $\frac{p_{j}^{se}}{p_{j+1}^{se}}=\left(\frac{p_{j}^{e}}{p_{j+1}^{e}}\right)^{\gamma}$
for $1\leq j\leq d_{e}-1$ and some constant $\gamma<1$, or (ii)
$d_{e}=3$. 

We present the proof for condition (i) in \cite{Note1}.
If this condition is fulfilled, the permutation $\mathcal{V}_{se}$
minimizes the dissipated heat. The corresponding environment state
$\sigma_{e}$ is a thermal state of temperature $\gamma\beta_{e}<\beta_{e}$
\cite{Note1}, where $\beta_{e}$ is the inverse temperature
of $\rho_{e}$. The example used to illustrate condition (i) of Theorem
2 also satisfies (i) of Theorem 3 if $p_{1}^{s}/p_{2}^{s}=e^{\gamma\beta_{e}\omega_{e}}$,
with $\gamma=1/2$ \cite{Note1}. On the other hand, when
condition (i) does not hold but (ii) does, we can still achieve minimium
heat dissipation \textit{using a catalyst}, as we show below. This
catalytic transformation is of the form $\sigma_{e}\otimes\rho_{v}\rightarrow\rho'_{ev}$,
being $\rho'_{e}=\textrm{Tr}_{s}\rho'_{ev}$ a thermal state having
the same entropy $\sigma_{e}$. Therefore, the total transformation
$\rho_{se}\rightarrow\sigma_{s}\otimes\sigma_{e}\rightarrow\sigma_{e}\otimes\rho_{v}\rightarrow\rho'_{ev}$
also implements optimal erasure. We also remark that, if $d_{e}=3$,
$\rho_{se}$ can obey Theorem 2 only through the associated condition
(ii). 

\textit{Proof for condition (ii) of Theorem 3}. By applying the protocol
introduced in \cite {sparaciari2017energetic}, one can use an infinite
catalyst to transform $\sigma_{e}$ into the thermal state $\rho'_{e}$
that has the same entropy of $\sigma_{e}$, i.e. $S(\rho'_{e})=S(\rho_{e})-\Delta S_{s}$.
This is always possible by first applying a local unitary $U_{e}$
that yields a passive state $U_{e}\sigma_{e}U_{e}^{\dagger}$, which
provides the initial state for the aforementioned protocol \cite {sparaciari2017energetic}.
The resulting state $\rho'_{e}=\rho_{e}^{th}[S(\rho_{e})-\Delta S_{s}]$
yields the minimum dissipated heat for entropy erasure $-\Delta S_{s}$. 

\textbf{\textit{Illustrative example}}. We illustrate the prediction of Theorem 1 for catalytic mitigation of dissipation via correlations, by considering erasure of a maximally mixed qubit coupled to a harmonic oscillator (HO). The mitigation is numerically optimized with respect to the catalyst dimension $d_{v}$. We also emphasize that this example
is not based on the optimal transformations characterized in Theorems
2 and 3. Instead, for a HO initialized in a thermal state $\rho_{e}=e^{-\beta_{e}H_{e}}/\textrm{Tr}\left(e^{-\beta_{e}H_{e}}\right)$,
erasure is performed through the paradigmatic Jaynes-Cummings interaction
$H_{int}=|0_{s}\rangle\langle1_{s}|\otimes a_{e}^{\dagger}+|1_{s}\rangle\langle0_{s}|\otimes a_{e}$,
where $a_{e}$ is the anihilation operator of the HO. For simplicity,
we set the interaction strenght equal to 1. 

The total $se$ Hamiltonian reads $H_{se}=H_{s}+H_{e}+H_{int}$, where
$H_{e}=\omega_{e}a_{e}^{\dagger}a_{e}=\sum_{j=1}^{\infty}j\omega_{e}|j_{e}\rangle\langle j_{e}|$,
$H_{s}=\omega_{e}|1_{s}\rangle\langle1_{s}|$, and $\omega_{e}$ is
the transition frequency of $e$. Importantly, we can derive the exact
state $\sigma_{se}$ since the evolution $U_{se}=e^{-iH_{se}t}$ is
analytically solvable. The catalytic transformation is characterized by
permutations $\mathcal{U}_{sev}^{(1)}=|1_{s}\rangle\langle1_{s}|\otimes\bigoplus_{k=1}^{d_{v}-1}\mathcal{U}_{2_{e}k_{v}\leftrightarrow1_{e}(k+1)_{v}}$
and $\mathcal{U}_{sev}^{(2)}=|2_{s}\rangle\langle2_{s}|\otimes\mathcal{U}_{2_{e}1_{v}\leftrightarrow1_{e}d_{v}}$.
The resulting unitary $U_{se}=\mathcal{U}_{sev}^{(1)}\oplus\mathcal{U}_{sev}^{(2)}$
has the structure of the general unitaries constructed for the sufficiency
proof of Theorem 1, with $(I,I',J,J')=(2,1,1,2)$. Accordingly, $U_{se}$
only couples the levels $|1_{e}\rangle$ and $|2_{e}\rangle$ of the
environment, and Fig. 1(b) illustrates its effect for the case $d_{v}=3$. 

\begin{figure}

\centering{}\includegraphics[scale=0.37]{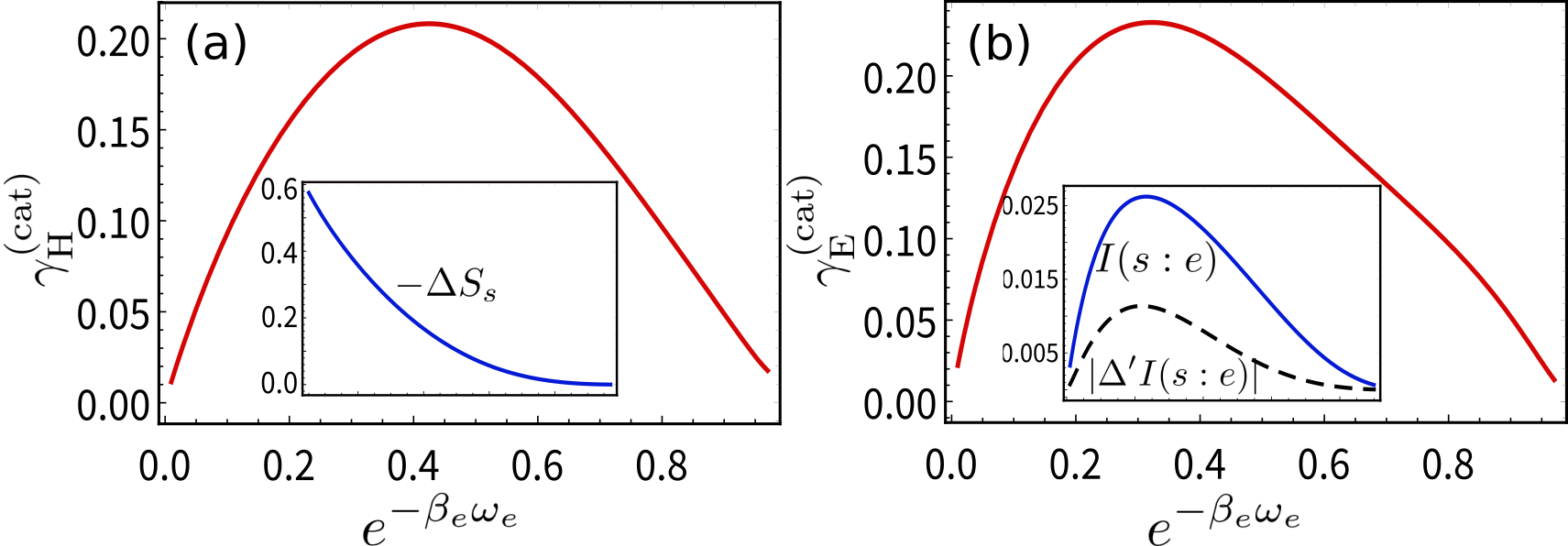}\caption{Catalytic performance coefficients $\gamma_{\textrm{H}}^{(\textrm{cat})}$ (a) and $\gamma_{\textrm{E}}^{(\textrm{cat})}$ (b), for the erasure process described in the main text. These coefficients quantify how effective is the catalyst at mitigating heat waste [Eq. (\ref{eq:3 gamma_h})] or entropy waste [Eq. (\ref{eq:4 gamma_e})]. Horizontal axes in the insets are identical to horizontal axes in the main plots. $\omega_{e}$ and $\beta_{e}$ are respectively the frequency and initial inverse temperature of a harmonic oscillator, which is utilized to erase entropy $-\Delta S_{s}$ (inset of (a)) from a maximally mixed qubit. The inset of (b) shows the mutual information generated after erasure (blue solid), and the mutual information that the catalyst consumes in the mitigation task (black dashed). }
\label{Fig. 2}
\end{figure}

To assess the performance of the catalyst, we introduce the coefficients 

\begin{align}\label{eq:3 gamma_h}
	\gamma_{\textrm{H}}^{(\textrm{cat})}&	=\frac{Q_{e}-Q'_{e}}{Q_{e}+T_{e}\Delta S_{s}}, \\
	\label{eq:4 gamma_e}\gamma_{\textrm{E}}^{(\textrm{cat})}&	=\frac{-\Delta'S_{e}}{\Delta S_{e}+\Delta S_{s}}=\frac{-\Delta'S_{e}}{I(s:e)}.
\end{align}
The coefficient $\gamma_{\textrm{H}}^{(\textrm{cat})}$ evaluates how effective is the catalyst at mitigating heat dissipation. We recall that $Q_{e}$ is the heat dissipated by the erasure transformation, and $Q'_{e}$ is the total heat dissipated after using the catalyst. Moreover, $-T_{e}\Delta S_{s}$ is the theoretical minimum, corresponding to Landauer's limit. Hence, $0\leq\gamma_{\textrm{H}}^{(\textrm{cat})}\leq1$, and $\gamma_{\textrm{H}}^{(\textrm{cat})}=1$ only if $Q'_{e}=-T_{e}\Delta S_{s}$. Remarkably, Fig. \ref{Fig. 2}(a) shows that the catalyst achieves up to $\sim20\%$ of the maximum mitigation $Q_{e}+T_{e}\Delta S_{s}$, for $e^{-\beta_{e}\omega_{e}}\sim0.42$.  

A similar behavior is seen in Fig. \ref{Fig. 2}(b) for mitigation of entropy dissipation, quantified by the coefficient $\gamma_{\textrm{E}}^{(\textrm{cat})}$ [Eq.(\ref{eq:4 gamma_e})]. The numerator of Eq. (\ref{eq:4 gamma_e}) is the difference between the initially dissipated entropy, $\Delta S_{e}$, and the total entropy dissipated after using the catalyst, given by $\Delta S_{e}+\Delta'S_{e}$. The denominator stands for the maximum of this quantity, corresponding to $\Delta S_{e}+\Delta'S_{e}=-\Delta S_{s}$.

Interestingly, Eq. (\ref{eq:4 gamma_e}) can be recast in terms of the initial mutual information $I(s:e)$, which expresses the fact that the ``fuel'' consumed by the catalyst comes from the correlations generated after erasure. In particular, one can show that $\gamma_{e}^{(\textrm{cat})}=1$ requires $\Delta'I(s:e)=-I(s:e)$, i.e. that all the aforementioned correlations are catalytically consumed. In contrast, the inset of Fig. \ref{Fig. 2}(b) shows that $|\Delta'I(s:e)|<I(s:e)$ in general. 

\textbf{\textit{Conclusions}}. We showed that correlations generated in information erasure can be catalytically leveraged to mitigate the dissipation of heat and entropy into the environment. While maximum erasure with minimum dissipation is possible under special conditions, as we showed, any realistic erasure transformation is expected to produce correlations. The  catalytic advantage was illustrated in an erasure process performed via the well known Jaynes-Cummings interaction, showing a significant reduction of heat and entropy waste.  

In view that traditional thermodynamic bounds usually give loose predictions when applied to finite environments
\cite{reeb2014improved,uzdin2021passivity}, new and tighter thermodynamic limits can emerge when catalysts are combined with these environments. Our findings pave the way for further investigations along this line. Beyond information erasure, catalysts could be useful to mitigate dissipation in heat engines and refrigerators, thus improving the performance of these devices and leading to new efficiency bounds.
\vskip 2 mm
\begin{acknowledgments}
We thank Matteo Lostaglio for his careful reading
and feedback on a earlier version of this manuscript.
\end{acknowledgments}

\bibliographystyle{apsrev}
\bibliography{References.bib}

\onecolumngrid

\section*{Supplemental Material}

\setcounter{section}{0}
\global\long\def\thefigure{S\arabic{figure}}
\setcounter{figure}{0}
\global\long\def\theequation{S\arabic{equation}}
\setcounter{equation}{0}

\section{Majorization and dissipation in information erasure }

In information erasure dissipation may have two related meanings.
Usually, it refers to the heat released into the environment as a
consequence of the erasure process. This kind of dissipation is energetic
in nature and it is one of the fundamental quantities entering Landauer's
principle {[}Eq. (1) in the main text{]}. On the other hand, there
is also an information-like form of dissipation. This refers to the
increment of entropy in the environment, which follows from the entropy
reduction of the system and the generation of correlations between
both. 

We address the task of mitigating dissipation by considering the two
aspects described above. To this end, the notion of majorization \cite{marshall1979inequalities,nielsen2002introduction}
turns out to be specially suited. Using the notation of the main text,
$\sigma_{e}$ is the environment state after information erasure,
and $\rho'_{e}$ the corresponding final state. As we show below,
heat and entropy can simultaneously be reduced in transformations
$\sigma_{e}\rightarrow\rho'_{e}$ such that $\rho'_{e}$ majorizes
$\sigma_{e}$. Majorization is defined in terms of the eigenvalues
of $\rho'_{e}$ and $\sigma_{e}$, which we write as $\{p_{i}^{\prime e}\}_{j=1}^{d_{e}}$
and $\{q_{i}^{e}\}_{j=1}^{d_{e}}$, respectively. Without loss of
generality, we can assume that $p_{j}^{\prime e}\geq p_{j+1}^{\prime e}$
and $q_{j}^{e}\geq q_{j+1}^{e}$ for $1\leq j\leq d_{e}-1$. The state
$\rho'_{e}$ majorizes the state $\sigma_{e}$ (equivalently expressed
as $\rho'_{e}\succ\sigma_{e}$) if and only if $\sum_{j=1}^{J}p_{j}^{\prime e}\geq\sum_{j=1}^{J}q_{j}^{e}$,
for $1\leq J\leq d_{e}$. 

Let us consider first the mitigation of entropy dissipation through
majorization $\rho'_{e}\succ\sigma_{e}$. Since the von Neumann entropy
is a Schur-concave function, we have that $S(\rho'_{e})\leq
S(\sigma_{e})$.

The mitigation of heat dissipation is addressed through Lemma 1. We
remark that Eq. (\ref{eq:3 Lemma 1})  relates the energies
of passive states \cite{pusz1978passive,allahverdyan2004maximal} obtained from $\rho'_{e}$
and $\sigma_{e}$. This is in accordance with the viewpoint that
heat is an energy transfer which involves an \textit{entropy change},
while any remnant energy that can be unitarily extracted constitutes
work \cite{vinjanampathy2016quantum,niedenzu2018quantum}. Therefore, a proper evaluation of the heat dissipated
in the transformations $\rho_{e}\rightarrow\sigma_{e}$ and $\sigma_{e}\rightarrow\rho'_{e}$
requires to remove the work-like contribution to the energies $\textrm{Tr}\left(H_{e}\sigma_{e}\right)$
and $\textrm{Tr}\left(H_{e}\rho'_{e}\right)$, by transforming $\sigma_{e}$
and $\rho'_{e}$ into passive states. In this way, the mitigation
of heat dissipation in the transformation $\sigma_{e}\rightarrow\rho'_{e}$
is a direct consequence of Lemma 1. 

\textbf{Lemma 1}. For $\sigma_{e}$ and $\rho'_{e}$ environment states
such $\rho'_{e}\succ\sigma{}_{e}$, 
\begin{equation}
\textrm{min}_{U_{e}}\textrm{Tr}\left(H_{e}U_{e}\rho'_{e}U_{e}^{\dagger}\right)\leq\textrm{min}_{U_{e}}\textrm{Tr}\left(H_{e}U_{e}\sigma{}_{e}U_{e}^{\dagger}\right),\label{eq:3 Lemma 1}
\end{equation}
where $H_{e}$ is the Hamiltonian of the environment and $U_{e}$
is a unitary acting on it. 

\textit{Proof}. For a Hamiltonian with eigendecomposition $H_{e}=\sum_{j}\varepsilon_{j}^{e}|j_{e}\rangle\langle j_{e}|$,
let $\varrho_{e}=\sum_{j}r_{j}^{e}|j_{e}\rangle\langle j_{e}|$ and
$\varrho'_{e}=\sum_{j}r_{j}^{\prime e}|j_{e}\rangle\langle j_{e}|$
denote the passive states that minimize each side of Eq. (\ref{eq:3 Lemma 1}).
Since these states are obtained via unitary transformations, the eigenvalues
$\{r_{j}^{e}\}$ and $\{r_{j}^{\prime e}\}$ are the same eigenvalues
of $\rho_{e}$ and $\rho'_{e}$, respectively. Hence, $\rho'_{e}\succ\sigma{}_{e}$
is equivalent to $\varrho'_{e}\succ\varrho_{e}$, if and only if $\sum_{j=1}^{J}\left(r_{j}^{\prime e}\right)^{\downarrow}\geq\sum_{j=1}^{J}\left(r_{j}^{e}\right)^{\downarrow}$
for $1\leq J\leq d_{e}$. Assuming $\varepsilon_{j}^{e}\leq\varepsilon_{j+1}^{e}$,
passivity also implies that $r_{j}^{e}\geq r_{j+1}^{e}$, $r_{j}^{\prime e}\geq r_{j+1}^{\prime e}$,
and therefore $\left(r_{j}^{e}\right)^{\downarrow}=r_{j}^{e}$
and $\left(r_{j}^{\prime e}\right)^{\downarrow}=r_{j}^{\prime e}$. 

Given that the majorization relation is expressed via the partial
sums $S_{J}^{e}\equiv\sum_{j=1}^{J}r_{j}^{e}$ and $S_{J}^{\prime e}\equiv\sum_{j=1}^{J}r_{j}^{\prime e}$,
for our purpose it is convenient to write the average energies $\bigl\langle H_{e}\bigr\rangle$
in terms of these partial sums. For the state $\varrho_{e}$, we will
show that 

\begin{equation}
\textrm{Tr}\left(H_{e}\varrho_{e}\right)=-\sum_{J=1}^{d_{e}-1}S_{J}^{e}\left(\varepsilon_{J+1}^{e}-\varepsilon_{J}^{e}\right)+\varepsilon_{d_{e}}^{e}.\label{eq:4}
\end{equation}
In this way, we immediately see that $S_{J}^{\prime e}\geq S_{J}^{e}$
for $1\leq J\leq d_{e}$ (majorization $\varrho'_{e}\succ\varrho_{e}$)
implies $\textrm{Tr}\left(H_{e}\varrho'_{e}\right)=-\sum_{J=1}^{d_{e}-1}S_{J}^{\prime e}\left(\varepsilon_{J+1}^{e}-\varepsilon_{J}^{e}\right)\leq\textrm{Tr}\left(H_{e}\varrho_{e}\right)$.

First, note that $\sum_{J=1}^{1}S_{J}^{e}\left(\varepsilon_{J+1}^{e}-\varepsilon_{J}^{e}\right)=-r_{1}^{e}\varepsilon_{1}^{e}+S_{1}^{e}\varepsilon_{2}^{e}$
and $\sum_{J=1}^{2}S_{J}^{e}\left(\varepsilon_{J+1}^{e}-\varepsilon_{J}^{e}\right)=-r_{1}^{e}\varepsilon_{1}^{e}-r_{2}^{e}\varepsilon_{2}^{e}+S_{2}^{e}\varepsilon_{3}^{e}$.
This suggests the general formula 
\begin{equation}
\sum_{J=1}^{\mathcal{J}}S_{J}^{e}\left(\varepsilon_{J+1}^{e}-\varepsilon_{J}^{e}\right)=-\sum_{j=1}^{\mathcal{J}}r_{j}^{e}\varepsilon_{j}^{e}+S_{\mathcal{J}}^{e}\varepsilon_{\mathcal{J}+1}^{e},\label{eq:5}
\end{equation}
for $1\leq\mathcal{J}\leq d_{e}-1$. Assuming that it holds for $\mathcal{J}$,
for $\mathcal{J}+1$ we have: 
\begin{align}
\sum_{J=1}^{\mathcal{J}+1}S_{J}^{e}\left(\varepsilon_{J+1}^{e}-\varepsilon_{J}^{e}\right) & =-\sum_{j=1}^{\mathcal{J}}p_{j}^{e}\varepsilon_{j}^{e}+S_{\mathcal{J}}^{e}\varepsilon_{\mathcal{J}+1}^{e}+S_{\mathcal{J}+1}^{e}\left(\varepsilon_{\mathcal{J}+2}^{e}-\varepsilon_{\mathcal{J}+1}^{e}\right)\nonumber \\
 & =-\sum_{j=1}^{\mathcal{J}}p_{j}^{e}\varepsilon_{j}^{e}+\left(S_{\mathcal{J}}^{e}-S_{\mathcal{J}+1}^{e}\right)\varepsilon_{\mathcal{J}+1}^{e}+S_{\mathcal{J}+1}^{e}\varepsilon_{\mathcal{J}+2}^{e}\nonumber \\
 & =-\sum_{j=1}^{\mathcal{J}+1}p_{j}^{e}\varepsilon_{j}^{e}+S_{\mathcal{J}+1}^{e}\varepsilon_{\mathcal{J}+2}^{e}.\label{eq:6}
\end{align}
Therefore, we conclude by induction that Eq. (\ref{eq:5}) is valid
for $1\leq\mathcal{J}\leq d_{e}-1$. Equation (\ref{eq:4}) follows
by applying Eq. (\ref{eq:5}) with $\mathcal{J}=d_{e}-1$ and using
$S_{d_{e}-1}^{e}+p_{d_{e}}^{e}=1$.

\section{Lemmas utilised for the proof of theorem 1}

\textbf{Lemma 2}. Let $\sigma_{se}=\sum_{i=1}^{d_{s}}\sum_{j=1}^{d_{e}}q_{i,j}^{se}|i_{s}j_{e}\rangle\langle i_{s}j_{e}|$
describe a \textit{classical} state of the system and the environment,
having eigenvalues $\{q_{i,j}^{se}\}_{1\leq i\leq d_{s},1\leq j\leq d_{e}}$.
If $\sigma_{se}$ is correlated, $\frac{q_{I,J}^{se}}{q_{I,J'}^{se}}>\frac{q_{I',J}^{se}}{q_{I',J'}^{se}}$
for some tuple $(I,I',J,J')$. 

\textit{Proof}. Let us first express the inequality $\frac{q_{I,J}^{se}}{q_{I,J'}^{se}}>\frac{q_{I',J}^{se}}{q_{I',J'}^{se}}$
in the equivalent form $\frac{q_{I,J}^{se}}{q_{I',J}^{se}}>\frac{q_{I,J'}^{se}}{q_{I',J'}^{se}}$.
We show that a correlated state $\sigma_{se}$ implies this inequality
(for at least one $(I,I',J,J')$), by proving that if 
\begin{equation}
{\normalcolor {\normalcolor {\color{red}{\normalcolor \frac{q_{I,J}^{se}}{q_{I',J}^{se}}=\frac{q_{I,J'}^{se}}{q_{I',J'}^{se}}\textrm{ for all }(I,I',J,J')}}}}\label{eq:7}
\end{equation}
then $\sigma_{se}$ is uncorrelated. Defining ${\color{red}{\normalcolor Q_{I,K}^{se}\equiv\sum_{k=1}^{K}q_{I,k}^{se}}}$,
we will apply an inductive strategy to prove this claim. 

Specifically, we start by inductively deriving the identity 
\begin{equation}
\frac{Q_{I,K}^{se}}{Q_{I',K}^{se}}=\frac{q_{I,K}^{se}}{q_{I',K}^{se}},\textrm{ for }0\leq K\leq d_{e}\label{eq:8}
\end{equation}
and $1\leq I,I'\leq d_{s}$. By definition of $Q_{I,K}^{se}$, for
$K=1$ (\ref{eq:8}) holds trivially. Since $\frac{q_{I,K}^{se}}{q_{I',K}^{se}}=\frac{q_{I,K+1}^{se}}{q_{I',K+1}^{se}}$,
according to (\ref{eq:7}), the validity of Eq. (\ref{eq:8}) for
$K$ is equivalent to $\frac{Q_{I,K}^{se}}{Q_{I',K}^{se}}=\frac{q_{I,K+1}^{se}}{q_{I',K+1}^{se}}$
for $1\leq I,I'\leq d_{s}$. Moreover, it is straightforward to check
that 
\begin{equation}
\frac{Q_{I,K}^{se}}{Q_{I',K}^{se}}=\frac{q_{I,K+1}^{se}}{q_{I',K+1}^{se}}\Leftrightarrow\frac{Q_{I,K+1}^{se}}{Q_{I',K+1}^{se}}\equiv\frac{Q_{I,K}^{se}+q_{I,K+1}^{se}}{Q_{I',K}^{se}+q_{I',K+1}^{se}}=\frac{q_{I,K+1}^{se}}{q_{I',K+1}^{se}},\label{eq:9}
\end{equation}
which shows that $\frac{Q_{I,K+1}^{se}}{Q_{I',K+1}^{se}}=\frac{q_{I,K+1}^{se}}{q_{I',K+1}^{se}}$
for $1\leq I,I'\leq d_{s}$, and completes the proof of Eq. (\ref{eq:8}). 

Since $\frac{q_{I,K}^{se}}{q_{I',K}^{se}}=\frac{q_{I,J}^{se}}{q_{I',J}^{se}}$
for $1\leq K,J\leq d_{e}$ (cf. (\ref{eq:7})), it follows from (\ref{eq:8})
that 
\begin{equation}
\frac{Q_{I,K}^{se}}{Q_{I',K}^{se}}=\frac{q_{I,J}^{se}}{q_{I',J}^{se}}\textrm{ for all }(I,I',K,J).\label{eq:10}
\end{equation}
In particular, 
\begin{equation}
\frac{Q_{I,d_{e}}^{se}}{Q_{I',d_{e}}^{se}}=\frac{q_{I,J}^{se}}{q_{I',J}^{se}}\textrm{ for }1\leq J\leq d_{e}\Leftrightarrow\frac{\sum_{j=1}^{d_{e}}q_{I,j}^{se}}{\sum_{j=1}^{d_{e}}q_{I',j}^{se}}=\frac{q_{I}^{s}}{q_{I'}^{s}}=\frac{q_{I,J}^{se}}{q_{I',J}^{se}}\textrm{ for }1\leq J\leq d_{e}.\label{eq:11}
\end{equation}

Using Bayes rule, $q_{I,J}^{se}=q_{I}^{s}q^{e}(J|I)$, where $q^{e}(J|I)$
denotes the conditional probability to measure $J$ in the environment
given that the system is measured in $I$, we can reexpress Eq. (\ref{eq:11})
as 
\begin{equation}
\frac{q_{I}^{s}}{q_{I'}^{s}}=\frac{q_{I}^{s}q^{e}(J|I)}{q_{I'}^{s}q^{e}(J|I')}\Leftrightarrow q^{e}(J|I)=q^{e}(J|I')\textrm{ for }1\leq J\leq d_{e}.\label{eq:12}
\end{equation}
Since Eq. (\ref{eq:12}) holds for all $I,I'$, it means that the
conditional probabilities $q^{e}(J|I)$ are independent of $I$. That
is, $q^{e}(J|I)=q^{e}(J)$ for all $I,J$, which leads us to conclude
that $\sigma_{se}$ must be uncorrelated. \\

\textbf{Remark 1}. The equivalence $\frac{q_{I,J}^{se}}{q_{I',J}^{se}}=\frac{q_{I,J'}^{se}}{q_{I',J'}^{se}}\Leftrightarrow\frac{q_{I,J}^{se}}{q_{I,J'}^{se}}=\frac{q_{I',J}^{se}}{q_{I',J'}^{se}}$
implies that the independence of the conditional probabilities $q^{e}(J|I)$
carries over the conditional probabilities $q^{s}(I|J)\equiv q_{I,J}^{se}/q_{J}^{e}$.
This is natural since a caracterization of (classical) correlations
in terms of conditional probabilities should be independent of the
system with respect to which such probabilities are defined. \\

\textbf{Lemma 3}. Consider the permutation $U_{sev}=\mathcal{U}_{sev}^{(I')}\oplus\mathcal{U}_{sev}^{(I)}$,
where $\mathcal{U}_{sev}^{(I')}=\bigoplus_{k=1}^{d_{v}-1}\mathcal{U}_{I'_{s}J'_{e}k_{v}\leftrightarrow I'_{s}J_{e}(k+1)_{v}}$
and $\mathcal{U}_{sev}^{(I)}=\mathcal{U}_{I_{s}J'_{e}1_{v}\leftrightarrow I_{s}J_{e}d_{v}}$
(cf. main text for details). As explained in the main text (proof
of the ``If'' implication of Theorem 1), each two-level permutation
$\mathcal{U}_{I'_{s}J'_{e}k_{v}\leftrightarrow I'_{s}J_{e}(k+1)_{v}}$
transfers population $\delta_{k}\equiv q_{I',J'}^{se}p_{k}^{v}-q_{I',J}^{se}p_{k+1}^{v}$
from $|k_{v}\rangle$ to $|(k+1)_{v}\rangle$, while $\mathcal{U}_{sev}^{(I)}$
transfers population $\delta_{d_{v}}\equiv q_{I,J}^{se}p_{d_{v}}^{v}-q_{I,J'}^{se}p_{1}^{v}$
from $|d_{v}\rangle$ to $|1_{v}\rangle$. 

For arbitrary eigenvalues $\{q_{i,j}^{se}\}$, there exist catalyst
eigenvalues $\{p_{k}^{v}\}_{k=1}^{d_{v}}$ such that 
\begin{equation}
\delta_{k}=\delta_{k+1}\textrm{ for }1\leq k\leq d_{v}-1.\label{eq:13}
\end{equation}

\textit{Proof}. Equations (\ref{eq:13}) are a set of $d_{v}-1$ linear
equations with $d_{v}$ unknowns $p_{k}^{v}$. Including the normalization
condition, $\sum_{k=1}^{d_{v}}p_{k}^{v}=1$, we have a total of $d_{v}$
linear equations. To solve them, we start by constructing a telescoping
sum that allows us to obtain a general formula for $p_{k+1}^{v}$,
in terms of $p_{1}^{v}$ and $p_{2}^{v}$. The cancellation of terms
in this sum follows from the expression 
\begin{equation}
\left(\frac{a}{b}\right)^{k-1}\left(bp_{k}^{v}-a{\color{red}{\normalcolor p_{k+1}^{v}}}\right)+\left(\frac{a}{b}\right)^{k}\left(b{\color{red}{\normalcolor p_{k+1}^{v}}}-ap_{k+2}^{v}\right)=\left(\frac{a}{b}\right)^{k-1}bp_{k}^{v}-\left(\frac{a}{b}\right)^{k}ap_{k+2}^{v},\label{eq:14}
\end{equation}
which is valid for arbitrary coefficients $a,b$. Therefore, setting
$a=q_{I',J}^{se}$ and $b=q_{I',J'}^{se}$, we obtain: 

\begin{align}
\sum_{k=1}^{K}\left(\frac{q_{I',J}^{se}}{q_{I',J'}^{se}}\right)^{k-1}\delta_{k} & =\sum_{k=1}^{K}\left(\frac{q_{I',J}^{se}}{q_{I',J'}^{se}}\right)^{k-1}\left(q_{I',J'}^{se}p_{k}^{v}-q_{I',J}^{se}p_{k+1}^{v}\right)\nonumber \\
 & =q_{I',J'}^{se}p_{1}^{v}-\left(\frac{q_{I',J}^{se}}{q_{I',J'}^{se}}\right)^{K-1}q_{I',J}^{se}p_{K+1}^{v}.\label{eq:15}
\end{align}

Since $\delta_{k}=\delta_{1}=q_{I',J'}^{se}p_{1}^{v}-q_{I',J}^{se}p_{2}^{v}$
for $1\leq k\leq d_{v}-1$ {[}cf. (\ref{eq:13}){]},
the sum $\sum_{k=1}^{K}\left(\frac{q_{I',J}^{se}}{q_{I',J'}^{se}}\right)^{k-1}\delta_{k}$
is also a geometric series that yields 
\begin{equation}
\sum_{k=1}^{K}\left(\frac{q_{I',J}^{se}}{q_{I',J'}^{se}}\right)^{k-1}\delta_{1}=\frac{1-\left(\frac{q_{I',J}^{se}}{q_{I',J'}^{se}}\right)^{K}}{1-\left(\frac{q_{I',J}^{se}}{q_{I',J'}^{se}}\right)}\delta_{1}.\label{eq:16}
\end{equation}
By combining Eqs. (\ref{eq:15}) and (\ref{eq:16}) and we find that:
\begin{equation}
p_{K+1}^{v}=\left(\frac{q_{I',J'}^{se}}{q_{I',J}^{se}}\right)^{K-1}\left[\left(\frac{q_{I',J'}^{se}}{q_{I',J}^{se}}\right)p_{1}^{v}-\frac{1}{q_{I',J}^{se}}\frac{1-\left(\frac{q_{I',J}^{se}}{q_{I',J'}^{se}}\right)^{K}}{1-\left(\frac{q_{I',J}^{se}}{q_{I',J'}^{se}}\right)}\delta_{1}\right].\label{eq:17}
\end{equation}
This equation gives $p_{K+1}^{v}$ in terms of $p_{1}^{v}$ and $p_{2}^{v}$
(and the coefficients $q_{I',J'}^{se},q_{I',J}^{se}$), for $1\leq K\leq d_{v}-1$. 

Next, we can use normalization $p_{1}^{v}+\sum_{K=1}^{d_{v}-1}p_{K+1}^{v}=1$
and (\ref{eq:17}) to find $p_{2}^{v}$ as a function of $p_{1}^{v}$.
This leaves all the $p_{K+1}^{v}$ depending solely on $p_{1}^{v}.$
Finally, the remaining unknown $p_{1}^{v}$ is derived by using the
equation in (\ref{eq:13}) corresponding to $k=d_{v}-1$, namely 
\begin{equation}
\delta_{d_{v}-1}=\delta_{d_{v}}\Leftrightarrow q_{I',J'}^{se}p_{d_{v}-1}^{v}-q_{I',J}^{se}p_{d_{v}}^{v}=q_{I,J}^{se}p_{d_{v}}^{v}-q_{I,J'}^{se}p_{1}^{v}.\label{eq:18}
\end{equation}
By applying Eq. (\ref{eq:17}) with $K=d_{v}-2$ and $K=d_{v}-1$,
we obtain $p_{d_{v}-1}^{v}$ and $p_{d_{v}}^{v}$ in terms of $p_{1}^{v}$,
respectively. The substitution of the resulting expressions into Eq.
(\ref{eq:18}) yields a linear equation in $p_{1}^{v}$ that can be
easily solved for this eigenvalue. \\

\textbf{Lemma 4}. Consider two states $\sigma_{e}=\sum_{j=1}^{d_{e}}q_{j}^{e}|j_{e}\rangle\langle j_{e}|$
and $\rho'_{e}=\sum_{j=1}^{d_{e}}p_{j}^{\prime e}|j_{e}\rangle\langle j_{e}|$,
such that $q_{J}^{e}>q_{J'}^{e}$, $p_{j}^{\prime e}=q_{j}^{e}$ for
$j\neq J,J'$, and $p_{J}^{\prime e}>q_{J}^{e}$. Then $\rho'_{e}\succ\sigma_{e}$. 

\textit{Proof}. The claim follows by directly applying the definition
of majorization. Once again, without loss of generality we can assume
that $\{q_{j}^{e}\}$ and $\{p_{j}^{\prime e}\}$ are sorted in decreasing
order, i.e. $q_{j}^{e}\geq q_{j+1}^{e}$ and $p_{j}^{\prime e}\geq p_{j+1}^{\prime e}$.
Therfore, for $q_{J}^{e}>q_{J'}^{e}$ we have that $J<J'$. Consider
all the partial sums $\sum_{j=1}^{K}q_{j}^{e}$ and $\sum_{j=1}^{K}p_{j}^{\prime e}$
such that $J\leq K<J'$. Since $p_{J}^{\prime e}>q_{J}^{e}$ and $p_{j}^{\prime e}=q_{j}^{e}$
if $j\neq J,J'$, we have that $\sum_{j=1}^{K}p_{j}^{\prime e}>\sum_{j=1}^{K}q_{j}^{e}$.
Moreover, $\sum_{j=1}^{K}p_{j}^{\prime e}=\sum_{j=1}^{K}q_{j}^{e}$
for $J'\leq K\leq d_{e}$. This implies that $\rho'_{e}\succ\sigma_{e}$. 

\section{Proof of theorem 2}

To prove Theorem 2 we shall resort to the state $\varrho_{e}$ defined
in Eq. (\ref{eq:20}). Moreover, we will use the subsets of indices
$K_{n}\equiv\{nd_{e}+1,nd_{e}+2,...,(n+1)d_{e}\}$, where $0\leq n\leq d_{s}-1$.
\\

\textbf{Remark 2}. Note that we can write $\{k\}_{k=1}^{d_{s}d_{e}}$
as $\{k\}_{k=1}^{d_{s}d_{e}}=\bigcup_{n=0}^{d_{s}-1}K_{n}$, and that
any $k'\in K_{n+1}$ can be obtained from some $k\in K_{n}$ as $k'=k+d_{e}$.\\

\textbf{Definition of }$\varrho_{e}$. Let $\rho_{se}=\rho_{s}\otimes\rho_{e}$
denote an uncorrelated state of $se$ with eigendecomposition $\rho_{se}=\sum_{k=1}^{d_{s}d_{e}}p_{k}^{se}|k_{se}\rangle\langle k_{se}|$,
$p_{k}^{se}\geq p_{k+1}^{se}$ for $1\leq k\leq d_{s}d_{e}-1$. Moreover,
suppose that the ratios $r_{k}^{se}\equiv\frac{p_{k}^{se}}{p_{k+1}^{se}}$
are periodicin $1\leq k\leq d_{e}-1$, with period
$d_{e}$. That is, $r_{k}^{se}=r_{k+\lambda d_{e}}^{se}=\frac{p_{k+\lambda d_{e}}^{se}}{p_{k+1+\lambda d_{e}}^{se}}$
for $1\leq k\leq d_{e}-1$ and $\lambda$ natural. We remark that
$k=d_{e}$ is not included, and thus we can have $r_{d_{e}}^{se}\neq r_{d_{e}+\lambda d_{e}}^{se}$.
For simplicity, we will not explicitly mention the domain $1\leq k\leq d_{e}-1$ when
referring to the condition $r_{k}^{se}=r_{k+\lambda d_{e}}^{se}$. 

Consider now $l\in K_{n}$ and $l'=l+d_{e}\in K_{n+1}$ (cf. Remark 2). Noting that for $1\leq k_{1},k_{2}\leq d_{e}$ any ratio $\frac{p_{k_{1}}^{se}}{p_{k_{2}}^{se}}=\prod_{k=k_{1}}^{k_{2}-1}r_{k}^{se}$
also has period $d_{e}$ (i.e. $\frac{p_{k_{1}+\lambda d_{e}}^{se}}{p_{k_{2}+\lambda d_{e}}^{se}}=\frac{p_{k_{1}}^{se}}{p_{k_{2}}^{se}}$), we have that for any $k\in K_{n}$ we can write $\frac{p_{k}^{se}}{p_{l}^{se}}$ as $\frac{p_{k}^{se}}{p_{l}^{se}}=\frac{p_{k+d_{e}}^{se}}{p_{l+d_{e}}^{se}}=\frac{p_{k'}^{se}}{p_{l'}^{se}}$, for  $k'=k+d_{e}\in K_{n+1}$. This implies that $\sum_{k\in K_{n}}\frac{p_{k}^{se}}{p_{l}^{se}}=\sum_{k'\in K_{n+1}}\frac{p_{k'}^{se}}{p_{l'}^{se}}$, and 

\begin{equation}
\frac{p_{l}^{se}}{\sum_{k\in K_{n}}p_{k}^{se}}=\frac{p_{l'}^{se}}{\sum_{k'\in K_{n+1}}p_{k'}^{se}}
\end{equation}

Accordingly, the state 

\begin{equation}
\varrho_{e}\equiv\frac{1}{\sum_{k\in K_{n}}p_{k}^{se}}\sum_{k\in K_{n}}p_{k}^{se}|k_{e}\rangle\langle k_{e}|\label{eq:20}
\end{equation}
is the same for $0\leq n\leq d_{s}-1$. \\

\textit{Proof of Theorem 2}. We divide the proof into two parts. First,
we will show that the periodicity condition $r_{k}^{se}=r_{k+\lambda d_{e}}^{se}$
implies that a maximum erasure transformation $\rho_{se}\rightarrow\sigma_{se}$
also yields an uncorrelated state $\sigma_{se}=\sigma_{s}\otimes\sigma_{e}$.
In the second part, we prove that Theorem 2 provides \textit{sufficient}  conditions
to have $r_{k}^{se}=r_{k+\lambda d_{e}}^{se}$. 

\textit{Part I}. The Schur-concavity of the von-Neumann entropy implies
that the minimum of $S(\sigma_{s})$ over joint unitaries $U_{se}$
is achieved if $\sigma_{s}=\textrm{Tr}_{e}\sigma_{se}\succ\textrm{Tr}_{e}\left(U_{se}\rho_{se}U_{se}^{\dagger}\right)$,
for any $U_{se}$. To derive a state $\sigma_{se}$
that satisfies this condition, we will apply the Schur-Horn theorem
\cite{horn1954doubly}. This theorem states that, for any hermitian
matrix, the vector formed by its diagonal entries is majorized by
the vector constructed with its eigenvalues. Clearly, any unitary
$U_{se}$ applied on $\rho_{se}$ produces a hermitian matrix $U_{se}\rho_{se}U_{se}^{\dagger}$
with eigenvalues $\{p_{k}^{se}\}$. If the marginal $\textrm{Tr}_{e}\left(U_{se}\rho_{se}U_{se}^{\dagger}\right)$
has eigendecomposition $\sum_{i=1}^{d_{s}}r_{i}^{s}|\phi_{s}^{(i)}\rangle\langle\phi_{s}^{(i)}|$,
for $1\leq I\leq d_{s}$ we have that 

\begin{equation}
\sum_{i=1}^{I}r_{i}^{s}=\sum_{i=1}^{I}\sum_{j=1}^{d_{e}}\langle\phi_{s}^{(i)}j_{e}|U_{se}\rho_{se}U_{se}^{\dagger}|\phi_{s}^{(i)}j_{e}\rangle\leq\sum_{k=1}^{Id_{e}}p_{k}^{se},\label{eq:21}
\end{equation}
where the inequality follows from the Schur-Horn theorem. 

We also remark that Eq. (\ref{eq:21}) holds regardless of the sorting
of the $\{r_{i}^{s}\}$. In particular, for decreasing eigenvalues
$r_{i}^{s}\geq r_{i+1}^{s}$, the left hand side of (\ref{eq:21})
is the partial sum of the largest $I$ eigenvalues of $\textrm{Tr}_{e}\left(U_{se}\rho_{se}U_{se}^{\dagger}\right)$.
If we can find a state $\sigma_{s}=\textrm{Tr}_{e}\sigma_{se}$ (where
$\sigma_{se}$ unitarily obtained from $\rho_{se}$), with eigenvalues
$\{q_{i}^{s}\}$ such that $q_{i}^{s}\geq q_{i+1}^{s}$ and $\sum_{i=1}^{I}q_{i}^{s}=\sum_{k=1}^{Id_{e}}p_{k}^{se}$,
Eq. (\ref{eq:21}) is equivalent to $\sigma_{s}\succ\textrm{Tr}_{e}\left(U_{se}\rho_{se}U_{se}^{\dagger}\right)$.
We can easily obtain this state via a permutation that rearranges
the eigenvalues of $\rho_{se}$ as follows. Keeping in mind that $\{|k_{se}\rangle\}=\{|i_{s}j_{e}\rangle\}$,
this permutation assigns the $d_{e}$ largest eigenvalues of $\rho_{se}$
to $\{|1_{s}j_{e}\rangle\}_{j}$, the next largest $d_{e}$ eigenvalues
of $\rho_{se}$ to $\{|2_{s}j_{e}\rangle\}_{j}$, and so forth. Using
$\{k\}_{k=1}^{d_{s}d_{e}}=\bigcup_{n=0}^{d_{s}-1}K_{n}$ (cf. Remark
2), the resulting transformation reads 
\begin{align}
\rho_{se}\rightarrow\sigma_{se} & =\sum_{i}|i_{s}\rangle\langle i_{s}|\otimes\sigma_{e}^{(i)}\nonumber \\
 & =\sum_{i}|i_{s}\rangle\langle i_{s}|\otimes\sum_{j\in K_{i}}p_{j}^{se}|j_{e}\rangle\langle j_{e}|\nonumber \\
 & =\sum_{i}q_{i}^{s}|i_{s}\rangle\langle i_{s}|\otimes\sum_{j\in K_{i}}\frac{p_{j}^{se}}{\left(\sum_{j\in K_{i}}p_{j}^{se}\right)}|j_{e}\rangle\langle j_{e}|\nonumber \\
 & =\sigma_{s}\otimes\varrho_{e},\label{eq:22}
\end{align}
where in the last line we apply the definition (\ref{eq:20}). 

Notice that Eq. (\ref{eq:22}) yields $q_{i}^{s}=\sum_{j\in K_{i}}p_{j}^{se}$.
Since the cardinality of each $K_{i}$ is $d_{e}$, it follows that
$\sum_{i=1}^{I}q_{i}^{s}=\sum_{i=1}^{I}\sum_{j\in K_{i}}p_{j}^{se}=\sum_{k=1}^{Id_{e}}p_{k}^{se}$.
As mentioned before, this implies that $\sigma_{s}\succ\textrm{Tr}_{e}\left(U_{se}\rho_{se}U_{se}^{\dagger}\right)$.
In addition, the condition $r_{k}^{se}=r_{k+\lambda d_{e}}^{se}$
guarantees that the states $\sum_{j\in K_{i}}\frac{p_{j}^{se}}{\left(\sum_{j\in K_{i}}p_{j}^{se}\right)}|j_{e}\rangle\langle j_{e}|$
in the third line of (\ref{eq:22}) are all identical to $\varrho_{e}$.
This allows us to obtain the maximum erasure transformation $\rho_{se}\rightarrow\sigma_{se}=\sigma_{s}\otimes\sigma_{e}$,
with $\sigma_{e}=\varrho_{e}$. 

\textit{Part II}. Now, we show that if the statement of Theorem 2
is satisfied then $r_{k}^{se}=r_{k+\lambda d_{e}}^{se}$.We also note that the explicit permutation $\mathcal{V}_{se}$ referred
to in Theorem 2 is given by Eq. (\ref{eq:22}). Common features
of the states described in this theorem are that $\rho_{x=s,e}$ are
full-rank states, and that $\frac{p_{j}^{e}}{p_{j+1}^{e}}\geq\frac{p_{1}^{s}}{p_{d_{s}}^{s}}$
for $1\leq j\leq d_{e}-1$. In fact, it is straightforward to check
that if any of the states $\rho_{x=s,e}$ is not full-rank then it
is impossible to have $r_{k}^{se}=r_{k+\lambda d_{e}}^{se}$. This
is because $\rho_{se}$ would not be full-rank and consequently at
least one $r_{k}^{se}$ would diverge. 

For the sake of convenience, we rewrite here conditions (i) and (ii)
of Theorem 2:

(i) $d_{e}=md_{s}$ and $r_{e,j}\equiv\frac{p_{j}^{e}}{p_{j+1}^{e}}$
is periodic with period $m$, i.e. $r_{e,j+\lambda m}=r_{e,j}$, $\lambda\in\mathbb{N}$.

(ii) $d_{s}=md_{e}$ and $r_{s,i}\equiv\frac{p_{i}^{s}}{p_{i+1}^{s}}$
is periodic with period $d_{e}$, i.e. $r_{s,i+\lambda d_{e}}=r_{s,i}$,
$\lambda\in\mathbb{N}$.

Let us start with the proof for (ii). We express the eigenvalues of
$\rho_{se}$ in the form 

\begin{equation}
\{p_{k}^{se}\}=\bigcup_{j=1}^{d_{e}}\{p_{i}^{s}p_{j}^{e}\}_{i=1}^{d_{s}},\label{eq:23}
\end{equation}
where eigenvalues within each subset $\{p_{i}^{s}p_{j}^{e}\}_{i=1}^{d_{s}}$
are decreasing, i.e. $p_{i}^{s}p_{j}^{e}\geq p_{i+1}^{s}p_{j}^{e}$.
Moreover, the inequalities $\frac{p_{j}^{e}}{p_{j+1}^{e}}\geq\frac{p_{1}^{s}}{p_{d_{s}}^{s}}$
entail that the minimum eigenvalue of $\{p_{i}^{s}p_{j}^{e}\}_{i=1}^{d_{s}}$
is larger or equal than the maximum eigenvalue of $\{p_{i}^{s}p_{j+1}^{e}\}_{i=1}^{d_{s}}$:
$p_{d_{s}}^{s}p_{j}^{e}>p_{1}^{s}p_{j+1}^{e}$. For $d_{s}=md_{e}$,
we can further decompose each $\{p_{i}^{s}p_{j}^{e}\}_{i=1}^{d_{s}}$
into $m$ subsets of $d_{e}$ eigenvalues. By indexing the $n$th
subset with the elements of $K_{n}$, we obtain $\{p_{i}^{s}p_{j}^{e}\}_{i=1}^{d_{s}}=\bigcup_{n=0}^{m-1}\{p_{i}^{s}p_{j}^{e}\}_{i\in K_{n}}$.
Therefore, 
\begin{align}
\{p_{k}^{se}\} & =\bigcup_{j=1}^{d_{e}}\bigcup_{n=0}^{m-1}\{p_{i}^{s}p_{j}^{e}\}_{i\in K_{n}}\nonumber \\
 & =\left(\bigcup_{n=0}^{m-1}\{p_{i}^{s}p_{1}^{e}\}_{i\in K_{n}}\right)\bigcup\left(\bigcup_{n=0}^{m-1}\{p_{i}^{s}p_{2}^{e}\}_{i\in K_{n}}\right)\bigcup...\bigcup\left(\bigcup_{n=0}^{m-1}\{p_{i}^{s}p_{d_{e}}^{e}\}_{i\in K_{n}}\right),\label{eq:24}\\
\bigcup_{n=0}^{m-1}\{p_{i}^{s}p_{j}^{e}\}_{i\in K_{n}} & =\{p_{i}^{s}p_{j}^{e}\}_{i\in K_{0}}\bigcup\{p_{i}^{s}p_{j}^{e}\}_{i\in K_{1}}\bigcup...\bigcup\{p_{i}^{s}p_{j}^{e}\}_{i\in K_{m-1}}.\label{eq:25}
\end{align}

The second line of Eq. (\ref{eq:24}) and Eq. (\ref{eq:25}) serve
to emphasize the decreasing order of the eigenvalues. Specifically,
eigenvalues within each subset are decreasing with respect to the
associated index, and eigenvalues in a subset located at  the left of a union operation are larger or equal than eigenvalues in a subset at the right. This
means that $\textrm{min}\left[\{p_{i}^{s}p_{j}^{e}\}_{i\in K_{n}}\right]\geq\textrm{max}\left[\{p_{i}^{s}p_{j}^{e}\}_{i\in K_{n+1}}\right]$,
regarding Eq. (\ref{eq:25}), and $\textrm{min}\left[\{p_{i}^{s}p_{j}^{e}\}_{i\in K_{m-1}}\right]\geq\textrm{max}\left[\{p_{i}^{s}p_{j+1}^{e}\}_{i\in K_{0}}\right]$,
regarding the second line of (\ref{eq:24}). 

According to Eq. (\ref{eq:24}), the ratio $\frac{p_{k}^{se}}{p_{k+1}^{se}}$
for elements $p_{k}^{se},p_{k+1}^{se}\in\{p_{i}^{s}p_{1}^{e}\}_{i\in K_{0}}$
reads $\frac{p_{k}^{se}}{p_{k+1}^{se}}=\frac{p_{i_{k}}^{s}}{p_{i_{k}+1}^{s}}$,
where $i_{k}\in K_{0}$ depends on $k$. Taking into account that
each $K_{n}$ has cardinality $d_{e}$, $p_{k+\lambda d_{e}}^{se},p_{k+1+\lambda d_{e}}^{se}\in\{p_{i}^{s}p_{j}^{e}\}_{i\in K_{n}}$
(with the pair $(j,n)$ being determined by $\lambda$) and we have
$\frac{p_{k+\lambda d_{e}}^{se}}{p_{k+1+\lambda d_{e}}^{se}}=\frac{p_{i_{k}+\lambda d_{e}}^{s}}{p_{i_{k}+1+\lambda d_{e}}^{s}}$.
This leads to the implication 
\begin{equation}
\frac{p_{i}^{s}}{p_{i+1}^{s}}=\frac{p_{i+\lambda d_{e}}^{s}}{p_{i+1+\lambda d_{e}}^{s}}\Rightarrow\frac{p_{k}^{se}}{p_{k+1}^{se}}=\frac{p_{k+\lambda d_{e}}^{se}}{p_{k+1+\lambda d_{e}}^{se}}\textrm{ iff }r_{k}^{se}=r_{k+\lambda d_{e}}^{se}.\label{eq:26}
\end{equation}

In the case of condition (i), since $d_{e}=md_{s}$, a subset $\{p_{k}^{se}\}_{k\in K_{n}}\subset\{p_{k}^{se}\}$
is composed of $m$ subsets $\{p_{i}^{s}p_{j}^{e}\}_{i=1}^{d_{s}}$.
This represents the opposite situation to that described by (\ref{eq:23})
and (\ref{eq:24}), which is a consequence of $d_{e}$ being now a
multiple of $d_{s}$ and not the other way around. Specifically, 
\begin{align}
\{p_{k}^{se}\} & =\bigcup_{n=0}^{d_{s}-1}\{p_{k}^{se}\}_{k\in K_{n}}\nonumber \\
 & =\bigcup_{n=0}^{d_{s}-1}\bigcup_{j=nm+1}^{(n+1)m}\{p_{i}^{s}p_{j}^{e}\}_{i=1}^{d_{s}},\label{eq:27}
\end{align}
where we stress again the decreasing order within subsets and among
elements of consecutive subsets. In particular, $\textrm{min}\left[\{p_{k}^{se}\}_{k\in K_{n}}\right]\geq\textrm{max}\left[\{p_{k}^{se}\}_{k\in K_{n+1}}\right]$,
for subsets in the first line of (\ref{eq:27}), and $\textrm{min}\left[\{p_{i}^{s}p_{j}^{e}\}_{i=1}^{d_{s}}\right]\geq\textrm{max}\left[\{p_{i}^{s}p_{j+1}^{e}\}_{i=1}^{d_{s}}\right]$,
for subsets in the second line. 
\begin{figure}

\centering{}\includegraphics{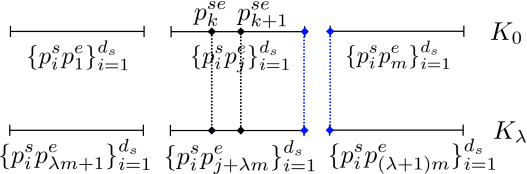}\caption{Visual guidance for Eqs. (\ref{eq:28})-(\ref{eq:30}). For simplicity,
we consider $m=3$. The argument is however valid for $m$ arbitrary. The upper segments represent the
set $K_{0}=\bigcup_{j=1}^{m}\{p_{i}^{s}p_{j}^{e}\}_{i=1}^{d_{s}}$,
and the lower segmets stand for $K_{\lambda}=\bigcup_{j=\lambda m+1}^{(\lambda+1)m}\{p_{i}^{s}p_{j}^{e}\}_{i=1}^{d_{s}}$.
For $p_{k}^{se},p_{k+1}^{se}\in\{p_{i}^{s}p_{j}^{e}\}_{i=1}^{d_{s}}$
(upper black dots), $p_{k+\lambda d_{e}}^{se},p_{k+1+\lambda d_{e}}^{se}$
correspond to the lower black dots. Since all the elements in the
upper middle segment possess the common factor $p_{j}^{e}$ and all
the elements in the lower middle segment possess the common factor
$p_{j+\lambda m}^{e}$, Eq. (\ref{eq:28}) follows. For $p_{k}^{se}\in\{p_{i}^{s}p_{j}^{e}\}_{i=1}^{d_{s}}$
(lefmost blue upper dot) and $p_{k+1}^{se}\in\{p_{i}^{s}p_{j+1}^{e}\}_{i=1}^{d_{s}}$
(rightmost blue upper dot), $p_{k+\lambda d_{e}}^{se},p_{k+1+\lambda d_{e}}^{se}$
are depicted by the corresponding lower blue dots. In this case $p_{k}^{se}=p_{d_{s}}^{s}p_{j}^{e}$
and $p_{k+1}^{se}=p_{1}^{s}p_{j+1}^{e}$, which leads to Eqs. (\ref{eq:29})
and (\ref{eq:30}).}
\label{Fig. 3}
\end{figure}

For eigenvalues $p_{k}^{se},p_{k+1}^{se}\in\{p_{k}^{se}\}_{k\in K_{0}}$
such that \textit{both} $p_{k}^{se}$ and $p_{k+1}^{se}$ belong to
$\{p_{i}^{s}p_{{\color{red}{\normalcolor j}}}^{e}\}_{i=1}^{d_{s}}\subseteq\{p_{k}^{se}\}_{k\in K_{0}}$,
Eq. (\ref{eq:27}) tells us that $\frac{p_{k}^{se}}{p_{k+1}^{se}}$
equals a ratio between two consecutive elements in $\{p_{i}^{s}\}$.
Moreover, $p_{k+\lambda d_{e}}^{se},p_{k+1+\lambda d_{e}}^{se}\in\{p_{i}^{s}p_{{\color{red}{\normalcolor j+\lambda m}}}^{e}\}_{i=1}^{d_{s}}\subseteq\{p_{k}^{se}\}_{k\in{\color{red}{\normalcolor K_{\lambda}}}}$
(see FIG. \ref{Fig. 3} and recall that each $K_{n}$ has cardinality $d_{e}$). Hence, 

\begin{equation}
\frac{p_{k}^{se}}{p_{k+1}^{se}}=\frac{p_{k+\lambda d_{e}}^{se}}{p_{k+1+\lambda d_{e}}^{se}}=\frac{p_{i_{k}}^{s}}{p_{i_{k}+1}^{s}},\label{eq:28}
\end{equation}
where $i_{k}$ depends on $k$. 

What remains is to show that the periodicity $r_{k}^{se}=r_{k+\lambda d_{e}}^{se}$
also holds if $p_{k}^{se}\in\{p_{i}^{s}p_{j}^{e}\}_{i=1}^{d_{s}}\subseteq\{p_{k}^{se}\}_{k\in K_{0}}$
and $p_{k+1}^{se}\in\{p_{i}^{s}p_{j+1}^{e}\}_{i=1}^{d_{s}}\subseteq\{p_{k}^{se}\}_{k\in K_{0}}$.
Note that due to the decreasing order this amounts to $p_{k}^{se}=p_{d_{s}}^{s}p_{j}^{e}$
and $p_{k+1}^{se}=p_{1}^{s}p_{j+1}^{e}$. Therefore, in this case
\begin{equation}
\frac{p_{k}^{se}}{p_{k+1}^{se}}=\frac{p_{d_{s}}^{s}}{p_{1}^{s}}\frac{p_{j_{k}}^{e}}{p_{j_{k}+1}^{e}},\label{eq:29}
\end{equation}
for some $j_{k}$ dependent on $k$. In addition, 
\begin{equation}
\frac{p_{k+\lambda d_{e}}^{se}}{p_{k+1+\lambda d_{e}}^{se}}=\frac{p_{d_{s}}^{s}}{p_{1}^{s}}\frac{p_{j_{k}+\lambda m}^{e}}{p_{j_{k}+1+\lambda m}^{e}},\label{eq:30}
\end{equation}
see also FIG. \ref{Fig. 3} .

From Eqs. (\ref{eq:29}) and (\ref{eq:30}) we have
the implication 
\begin{equation}
\frac{p_{j}^{e}}{p_{j+1}^{e}}=\frac{p_{j+\lambda m}^{e}}{p_{j+1+\lambda m}^{e}}\Rightarrow\frac{p_{k}^{se}}{p_{k+1}^{se}}=\frac{p_{k+\lambda d_{e}}^{se}}{p_{k+1+\lambda d_{e}}^{se}}\textrm{ iff }r_{k}^{se}=r_{k+\lambda d_{e}}^{se}.\label{eq:31}
\end{equation}
This concludes the proof corresponding to condition (i). 

\section{Proof of theorem 3 for condition (i)}

To begin with let us recall condition (i) of Theorem 3.

Let $\rho_{se}=\rho_{s}\otimes\rho_{e}$ be a state with marginals
$\rho_{x=s,e}$ that obey Theorem 2, and let $\rho_{se}=\sum_{j=1}^{d_{s}d_{e}}p_{j}^{se}|j_{se}\rangle\langle j_{se}|$
denote its eigendecomposition, with non-increasing eigenvalues $p_{j}^{se}\geq p_{j+1}^{se}$.
If $\rho_{e}$ is a thermal state, after the permutation $\mathcal{V}_{se}$
in Theorem 2 minimum heat dissipation is possible if (i) $\frac{p_{j}^{se}}{p_{j+1}^{se}}=\left(\frac{p_{j}^{e}}{p_{j+1}^{e}}\right)^{\gamma}$
for ${\color{red}{\normalcolor 1\leq j\leq}}{\color{red}{\normalcolor d_{e}-1}}$ and some constant $\gamma<1$. 

\textit{Proof}. In the proof of Theorem 2 we saw that $\mathcal{V}_{se}$
yields a transformation $\rho_{se}\rightarrow\sigma_{s}\otimes\sigma_{e}$,
where $\sigma_{e}=\varrho_{e}$ and $\varrho_{e}$ is given in Eq.
(\ref{eq:20}). Accordingly, 
\begin{equation}
\sigma_{e}=\frac{1}{\sum_{k\in K_{0}}p_{k}^{se}}\sum_{k\in K_{0}}p_{k}^{se}|k_{e}\rangle\langle k_{e}|=\frac{1}{\sum_{k=1}^{d_{e}}p_{k}^{se}}\sum_{k=1}^{d_{e}}p_{k}^{se}|k_{e}\rangle\langle k_{e}|.\label{eq:32}
\end{equation}
In this way, the quantities $\frac{p_{j}^{se}}{p_{j+1}^{se}}$ for
$1\leq j\leq d_{e}-1$ are the ratios between consecutive eigenvalues
of $\sigma_{e}$. It is also worth mentioning that since Theorem 2 implies the periodicity $\frac{p_{j}^{se}}{p_{j+1}^{se}}=\frac{p_{j+\lambda d_{e}}^{se}}{p_{j+1+\lambda d_{e}}^{se}}$, condition (i) of Theorem 3 is equivalent to $\frac{p_{j+\lambda d_{e}}^{se}}{p_{j+1+\lambda d_{e}}^{se}}=\left(\frac{p_{j}^{e}}{p_{j+1}^{e}}\right)^{\gamma}$, for $1\leq j\leq d_{e}-1$ and $\lambda\in\mathbb{N}$.

Given a Hamiltonian $H_{e}=\sum_{j=1}^{d_{e}}\varepsilon_{j}^{e}|j_{e}\rangle\langle j_{e}|$,
and a thermal state $\rho_{e}=\sum_{j=1}^{d_{e}}p_{j}^{e}|j_{e}\rangle\langle j_{e}|=\frac{e^{\beta_{e}H_{e}}}{\textrm{Tr}\left(e^{\beta_{e}H_{e}}\right)}$,
the condition $\frac{p_{j}^{se}}{p_{j+1}^{se}}=\left(\frac{p_{j}^{e}}{p_{j+1}^{e}}\right)^{\gamma}=e^{\gamma\beta_{e}(\varepsilon_{j+1}^{e}-\varepsilon_{j}^{e})}$
means that $\sigma_{e}$ is a thermal state at inverse temperature
$\gamma\beta_{e}<\beta_{e}$. This is the thermal state having entropy
$S_{e}-\Delta S_{s}$ (since $\rho_{se}\rightarrow\sigma_{s}\otimes\sigma_{e}$
dissipates minimum entropy), which minimizes the energy for such entropy
and consequently also minimizes the dissipated heat. \\

\textit{Example}. An example of state $\rho_{se}$ that adheres to
Theorem 2 and to condition (i) of Theorem 3 is the following. Consider
a two-level system $s$ and an environment $e$ of even dimension
with uniformly spaced energy spectrum. That is, $H_{e}=\sum_{j=1}^{d_{e}}j\omega_{e}|j_{e}\rangle\langle j_{e}|$.
At inverse temperature $\beta_{e}$, the ratios $\frac{p_{j}^{e}}{p_{j+1}^{e}}$
are given by the constant $\frac{p_{j}^{e}}{p_{j+1}^{e}}=e^{\beta_{e}\omega_{e}}$
and are therefore trivially periodic in $j$. Furthermore, $d_{e}=md_{s}=2d_{s}$.
If $e^{\beta_{e}\omega_{e}}>\frac{p_{1}^{s}}{p_{2}^{e}}$, we have
all the ingredients for the application of Theorem 2, with condition
(i). 

The inequalities $\frac{p_{j}^{e}}{p_{j+1}^{e}}\geq\frac{p_{1}^{s}}{p_{2}^{e}}$ give rise to decreasing eigenvalues $\{p_{j}^{se}\}=\{p_{1}^{s}p_{1}^{e},p_{2}^{s}p_{1}^{e},....,p_{1}^{s}p_{k}^{e},p_{2}^{s}p_{k}^{e},...\}$, for $1\leq k\leq d_{e}$. Therefore, 

\begin{align}
\frac{p_{j}^{se}}{p_{j+1}^{se}} & =\frac{p_{1}^{s}p_{(j+1)/2}^{e}}{p_{2}^{s}p_{(j+1)/2}^{e}}=\frac{p_{1}^{s}}{p_{2}^{s}}\textrm{ for }j\textrm{ odd},\label{eq:33}\\
\frac{p_{j}^{se}}{p_{j+1}^{se}} & =\frac{p_{2}^{s}p_{j/2}^{e}}{p_{1}^{s}p_{j/2+1}^{e}}\textrm{ for }j\textrm{ even}.\label{eq:34}
\end{align}
These equations imply that $\frac{p_{j}^{se}}{p_{j+1}^{se}}=\left(\frac{p_{j}^{e}}{p_{j+1}^{e}}\right)^{\gamma}$ iff $\frac{p_{1}^{s}}{p_{2}^{s}}=\frac{p_{2}^{s}p_{j/2}^{e}}{p_{1}^{s}p_{j/2+1}^{e}}=e^{\gamma\beta_{e}\omega_{e}}$, which is fulfilled for $\frac{p_{1}^{s}}{p_{2}^{s}}=e^{\gamma\beta_{e}\omega_{e}}$ with $\gamma=1/2$. In this way condition (i) of Theorem 3 is also satisfied.
In the particular case of a qubit environment ($d_{e}=2$), we have a single ratio $\frac{p_{j}^{e}}{p_{j+1}^{e}}=\frac{p_{1}^{e}}{p_{2}^{e}}$, which corresponds to $j=1$ odd. Therefore, only Eq. (\ref{eq:33}) is relevant and for $\frac{p_{1}^{s}}{p_{2}^{s}}=e^{\gamma\beta_{e}\omega_{e}}$ condition (i) of Theorem 3 holds, for \textit{any} $\gamma<1$.   

%\bibliographystyle{apsrev}
%\bibliography{References.bib}

\end{document}